\documentstyle[aps,multicol,epsfig]{revtex}
\newcommand{\beq}{\begin{equation}}
\newcommand{\eeq}{\end{equation}}
\newcommand{\bdis}{\begin{displaymath}}
\newcommand{\edis}{\end{displaymath}}
\newcommand{\bea}{\begin{eqnarray}}
\newcommand{\eea}{\end{eqnarray}}
\newcommand{\barr}{\begin{array}}
\newcommand{\earr}{\end{array}}
\title{Renormalization group approach to the critical behavior of
the forest fire model}
\author{V. Loreto$^{(1)}$, L. Pietronero$^{(1)}$, A. Vespignani$^{(2)}$ \\
and S. Zapperi$^{(3)}$}
\begin{document}
\maketitle
\centerline{1) {\em Dipartimento di Fisica, Universit\'a di Roma "La
Sapienza"}}
\centerline{{\em Piazzale Aldo Moro 2, 00185 Roma, Italy}}
\centerline{2) {\em Instituut-Lorentz, University of Leiden, P.O. Box 9506}}
\centerline{{\em 2300 RA, Leiden, The Netherlands}}
\centerline{3) {\em Center for Polymer Studies and Department of Physics}}
\centerline{{\em Boston University, Boston, MA 02215, USA}}
\date{~}
\medskip
\begin{abstract}
We introduce a Renormalization scheme for the one and two dimensional
Forest-Fire models in order to characterize the nature of the critical
state and its scale invariant dynamics.
We show the existence of a relevant scaling field associated with
a repulsive fixed point. This model is therefore critical in the usual sense
because the control parameter has to be tuned to its critical
value in order to get criticality. It turns out that this is not
just the condition for a time scale separation.
The critical exponents
are computed analytically and we obtain $\nu=1.0$, $\tau=1.0$ and
$\nu=0.65$, $\tau=1.16$ respectively for the one and two dimensional
case, in  very good agreement with numerical simulations.
\end{abstract}

PACS numbers: 64.60.Ak, 02.50.-r, 05.40.+j

%\newpage
\begin{multicols}{2}

\narrowtext

The Forest-Fire model (FFM) \cite{BAK1,DS} has been introduced as a possible
realization
of Self Organized Criticality (SOC) \cite{bak}.
The model formulated first by Bak and coworkers \cite{BAK1}
contains a tree growth probability $p$ and fire spreading to nearest
neighbours. In the limit of very slow tree growth, this model becomes
more and more deterministic \cite{GR1} and shows spiral-shaped
fire fronts. For this reason, the original model was modified by
Drossel and Schwabl \cite{DS} by introducing an ignition parameter $f$,
the lightning parameter. This parameter is the probability that during
one time step a tree without burning nearest neighbours becomes a burning tree.
In this case a critical behaviour, in the sense of
anomalous scaling laws, is observed in the double limit
$\theta=\frac{f}{p} \rightarrow 0$ and $p \to 0$.
These two limits describe a double separation of time scales: trees grow
fast compared to the occurrence of lightning in the system and forest
clusters burn down much faster than trees grow. This means that each
fire triggered by individual lightning does not overlap with other fires,
thus  clusters destroyed by fire are well defined objects. The
critical state is characterized by a power law distribution
$P(s) \sim s^{-\tau}$
of the forest clusters of $s$ sites and  the average  cluster radius scale as
$R \sim \theta^{-\nu}$. Since this critical state is reached independently
on the initial conditions and for a wide range of parameter values,
it is called self-organized critical state. This statement however is ambiguous
in that Self Organized Criticality refers to
the tendency of large dynamical systems to evolve {\em spontaneously} in a
critical state  without the fine tuning of any critical parameter.
 From this point of view, the role played by the parameter
$\theta$ is not clear.
In fact it seems an effective relevant parameter, as the reduced
temperature in thermal phase transitions, in that it allows criticality just
for its critical value $\theta=0$.

Usually the limit $\theta \rightarrow 0$ is referred to the existence of a
slow driving of the system or, in other words, a time scale separation, also
present in the definition of SOC models. It is worth to stress that in the SOC
models the requirement of a slow driving of the system is essentially
different.
In this case, in fact, the time scale separation just affects
the lower cut-off of the avalanche size distribution, the upper cut-off
being affected by finite-size effects \cite{CLP}.
In the FFM the parameter $\theta$, on the
contrary, affects directly the upper cut-off \cite{GR2}
and then it seems to play, in the
language of ordinary critical phenomena, the role of a relevant parameter.

In the context of Renormalization Group (RG) methods, a self-organized critical
phenomenon
can be viewed as a critical phenomenon in which all the parameters are
irrelevant, namely the scale invariant dynamics
corresponds to an attractive fixed point in the parameter space (phase space).
To address the study of intrinsically critical phenomena, we have developed
the Fixed Scale Transformation  approach to fractal growth and
recently we have introduced an RG scheme of novel type to study
sandpile models \cite{fst,PVZ1}.

In this letter we follow the same ideas in order to study the FFM
including the ignition parameter $f$.
This allows us  to clarify the role of the critical parameters $\theta$ and $p$
and
the nature of the critical state. In fact, we obtain that $\theta$
is the relevant control parameter,
so that FFM belongs more to the rich domain of
ordinary critical phenomena than to self-organized criticality. In addition we
are able to compute analytically the critical exponents characterizing the
model
in $d=1$ and $d=2$.

For the sake of clarity we recall briefly the dynamical rules of the model we
are going to discuss. At each time step the lattice is updated as follows:
{\bf i)} a burning tree becomes an empty site;
{\bf ii)} a green tree becomes a burning tree if at least one of its neighbors
is
burning; {\bf iii)} at an empty site a tree can grow with probability $p$;
{\bf iv)} a tree without burning nearest neighbors becomes a burning tree with
probability $f$.

Starting with arbitrary initial conditions, the system approaches, after a
short transient, a steady state the  properties of which depend on the
parameter  values.
Let $\rho_{0}$, $\rho_{1}$ and $\rho_{2}$ be the mean densities of empty sites,
of trees and of burning trees in the stationary state. These
probabilities describe in full generality the stationary properties of the
model,
while, as for the dynamical properties, we will refer to a phase space defined
by the parameters $p$ and $\theta=\frac{f}{p}$. In such a scheme $p$ and
$\theta$
parameterize the dynamics of the system.

We should  now extend the characterization of the stationary and
dynamical properties of the system at a generic scale $b^{(k)}=b_{0}2^{k}$ by
considering coarse-grained variables $p^{(k)}$, $f^{(k)}$ and
${\displaystyle \bf
\rho}^{(k)}=(\rho^{(k)}_{0},\rho^{(k)}_{1},\rho^{(k)}_{2})$.
In the following we will mainly refer to the two-dimensional case but all the
procedure can be drawn also in the one-dimensional case.
The stationary properties of coarse grained variables are defined as follows
(Fig.1): A cell of size $b^{(k)}$ is considered as "green" if it is spanned
from left to right by a connected path of green sites at scale
$b^{(k-1)}$. On the opposite a cell is empty if it is not spanned by a
connected
path of green sites. Finally, we consider a cell as burning if it contains at
least one burning tree. In the latter case the spanning condition is not
necessary because the fire spreads automatically to nearest neighbors sites.
Consequently, the dynamical properties of the coarse grained description of the
system will be characterized by the variables $p^{(k)}$ and
$\theta^{(k)}=\frac{f^{(k)}}{p^{(k)}}$.
The first variable express the probability that an empty cell of size $b^{(k)}$
becomes a green tree. The second one is defined using the probability $f^{(k)}$
that a green cell of size $b^{(k)}$ becomes a burning cell because of the
arrival
of a lightning event.

We now proceed to define a renormalization transformation for the dynamical
variables. We will use a cell-to-site transformation on the square lattice,
in which each cell at scale $b$ is formed by four sub-cells at scale $b/2$.
Every cell at the larger scale is then characterized by a number of green
and  empty sub-cells ranging from one to four.
The relative weight of each green configuration is
given by the probability to have the
corresponding number of green and empty sub-cells
\beq
W_{\alpha}({\displaystyle \bf \rho}) = n_{\alpha}\rho_0^{\alpha}
\rho_1^{4-\alpha},
\eeq
where $\alpha$ is  the number of empty sub-cells
and $n_{\alpha}$ is a normalization factor that takes into
account the multiplicity of the configurations.
In order to define a renormalization
transformation we start with an empty or green cell configuration at scale $b$
and study how it evolves using the dynamical rules of the model.
Here we consider a transformation defined by the following rules:
i) Every series of tree growth processes at scale $b/2$ that span
an empty cell at scale $b$ in the horizontal direction is renormalized in
the growth probability $p$ at scale $b$.
ii) Every lightning process at scale $b/2$ that affects a green cell of scale
$b$ contributes to the renormalization of the lightning probability $f$.
The spanning rule implies that only tree growth processes extending over
the size of the new length scale contribute to the renormalized dynamics.
Moreover, it ensures the connectivity properties of the green sites
in the  renormalization procedure.
An example of such a renormalization procedure for the lightning probability
is shown in fig.2 for the simplest calculation scheme; i.e. 2x2 cells with
left-right
spanning condition..
We have two possible processes
depending on whether both or just one green sub-cell is hit
by a lightning event.
The first case occurs with probability $2 f^{(k)}(1-f^{(k)})$,
the latter with probability $(f^{(k)})^2$.
By summing these probabilities one obtains, the renormalization equation
for the configuration with $\alpha=2$.
We can write a renormalization equation for
each configuration $\alpha$ corresponding to a green site, and,
averaging over all the
configurations we obtain:
\beq
f^{(k+1)}=\sum_{\alpha=2}^{4} W_{\alpha}f_{\alpha}^{(k+1)}
\label{fren1}
\eeq
with
\beq
\begin{array}{l}
f_{2}^{(k+1)}=f^{(k)}(2-f^{(k)})\\
f_{3}^{(k+1)}=f^{(k)}(3-3f^{(k)}+(f^{(k)})^2)\\
f_{4}^{(k+1)}=f^{(k)}(4-6f^{(k)}+4(f^{(k)})^2-(f^{(k)})^3)
\end{array}
\label{fren2}
\eeq
In an analogous way we can write the renormalization equations for $p^{(k+1)}$.
Let us consider an empty cell at scale $b$ of type $\alpha=2$ (see Fig.3).
After an updating
step, the cell will become green if at least one of the two empty sub-cells has
become green. This occurs with probability $(p^{(k)})^2$ and $2
p^{(k)}(1-p^{(k)})$,
corresponding to the fact that  both or only one sub-cell become green,
respectively.
In a similar way one can also write a
renormalization equation for each configuration $\alpha$
corresponding to an empty site,
and finally averaging over all the configurations we obtain
\beq
p^{(k+1)}=\sum_{\alpha=0}^{2}W_{\alpha}p_{\alpha}^{(k+1)}
\eeq
with
\beq
\begin{array}{l}
p_{0}^{(k+1)}=(p^{(k)})^2 (2-(p^{(k)})^2)\\
p_{1}^{(k+1)}=p^{(k)}(1+p^{(k)}-(p^{(k)})^2)\\
p_{2}^{(k+1)}=p^{(k)}(2-p^{(k)}),
\end{array}
\label{pren2}
\eeq
 From the previous equations it is then straightforward to obtain
the renormalization equation $\theta^{(k+1)} =f^{(k+1)}/p^{(k+1)}$.
These renormalization equations are not yet complete because the statistical
weights $W_\alpha$ are function of the the density vector ${\displaystyle
\bf\rho}^{k}$.
In fact, in order to describe the {\em stationary critical state} it is
necessary to couple
the dynamics  to a stationarity condition that gives the renormalization
equations
for the density vector. This scheme is similar to that used in \cite{PVZ1}.
The stationarity condition is obtained
from the master equations for the density vector in the mean field regime
by imposing the asymptotic equilibrium condition ($t \rightarrow \infty$)
\cite{DS1,CH},

\beq
\left\{
\begin{array}{l}
\rho_0^{(k)}=(1-\rho_1^{(k)})a^{(k)}/p^{(k)}\\
\rho_1^{(k)}=\frac{a^{(k)}}{\theta^{(k)}p^{(k)}+
4 \cdot a^{(k)}-a^{(k)}\cdot \rho_1^{(k)}\cdot {(2d-1)}}\\
\rho_2^{(k)}=(1-\rho_1^{(k)})a^{(k)}
\end{array}
\right .
\label{campomedio}
\eeq

where we defined $a^{(k)}=p^{(k)}/(1+p^{(k)})$,
with the normalization condition
$\rho_0^{(k)}+\rho_1^{(k)}+\rho_2^{(k)}=1$.
The stationarity condition, summarized in the (\ref{campomedio}),
provides the renormalized density vector at each scale, and it
couples the dynamical properties to the stationary ones.
It is worth to remark that we do not determine the RG equations for
${\displaystyle \bf \rho}$
from the coarse graining prescriptions.
In fact the stationary properties have to be evaluated considering
the average over many dynamical processes. Thus the densities
${\displaystyle \bf \rho}$ are determined from the renormalized
dynamical description of the system, namely  eq.s (\ref{campomedio})
with renormalized parameters.
Note also that the RG equations are written in the hypothesis of a
double time scale separation: fire spreading and trees growth are not
interacting. This is expressed by the consistency relation $p<1/T(\xi)$,
where $T(\xi)$ is the average time needed to burn a
cluster of characteristic size $\xi$. Since $T(\xi) \sim \xi^z$ where
$z$ is the  dynamical exponent, and $\xi \sim (f/p)^{-\nu}$ we have that
$T(\xi) \sim (f/ p)^{-\nu^\prime}$with $\nu^\prime=\nu z$.
Our RG approach is then suitable to describe what happens in
the region of the phase space $(p,\theta)$ defined by the condition
$p<\theta^{-\nu^\prime}$.

Given this scheme, we can find the fixed points of the
renormalization transformation by
studying the flow diagram
in the phase space of the parameters $({\displaystyle \bf \rho},p,\theta)$.
In two dimensions the RG equations show the fixed point $p^{*}=0$,
$\theta^{*}=0$
and ${\displaystyle \bf \rho}^{*}= (2/3,1/3,0)$.
A complete characterization of the fixed point
is obtained by the linearized RG equations around $p^{*}$,$\theta^{*}$ and
${\displaystyle \bf \rho}^{*}$. We find only one relevant scaling field that
corresponds to
an eigenvalue greater than 1, which is given by
\beq
\lambda=\frac{d \theta^{(k+1)}}{d\theta ^{(k)}}\mid_{p^{*},\theta^{*},
{\displaystyle \bf \rho}^{*}}
\eeq
Therefore, we have that the fixed point is repulsive in the $\theta$ direction,
which of course defines the relevant control parameter.
 Since the fixed point is repulsive we can determine the exponent
of the clusters characteristic length by using the largest eigenvalue, i.e.,
$\nu=\log 2 / \log \lambda$. We  obtain
$\nu=0.73$ by using the simple 2x2 cell renormalization
scheme. We can easily improve the results with a 3x3 cell calculation, which
gives
$\nu=0.65$, showing that the numerical result converges to the right value with
refined renormalization scheme.
The exponent $\nu$ describes the divergence of the correlation length by
$R \sim \theta^{-\nu}$ and the value obtained is in good agreement with the
value
$\nu \simeq 0.58$ measured in \cite{GR2,clar}.
In this perspective the parameter $\theta=\frac{f}{p}$ plays the role of the
relevant critical parameter as the reduced temperature in the thermal phase
transitions. For each value of $\theta$, small but finite, the system
shows an upper characteristic length in the cluster distribution. Only for
$\theta^*=0$ the system is critical and shows an infinite correlation length.

The exponent $\tau$ describing the distribution of fire spreading
can be obtained as follows.
The fires are
represented by the clusters of connected sites interested by a burning
process.
As in \cite{PVZ1} we define $K$  as
the probability that
an active relaxation process (i.e. fire) is limited between
the scales $b^{(k-1)}$ and $b^{(k)}$ and it does not extend further:

\beq
K=\int_{b^{(k-1)}}^{b^{(k)}}P(r)dr  \left/ \int_{b^{(k-1)}}^{\infty}P(r)dr
\right.
\label{K}
\eeq

where $P(r) dr$ is the probability to have a burning cluster with radius
between $r$ and $r+dr$.
In two dimensions with simple scaling arguments we can conclude that,
if $P(s) \sim s^{-\tau}$ and $s \sim r^{2}$ (compact clusters), then
$P(r) \sim r^{1-2\tau}$.
Inserting this expression in eq.(\ref{K}) we obtain (in $d=2$)
$\tau=1-(\log(1-K))/(2 \log 2)$.
In our case $K$ is the probability that at a generic scale $(k)$ all
the nearest neighbors of a burning tree are empty and then
$K=(1-\rho_{1}^{(k)})^{4}$.
In the scale invariant regime $(\rho_{1}^{(k)}=\rho_{1}^{*})$,
$K=0.1975$ and then we obtain for $\tau$ the value
$\tau \simeq 1.16$
in excellent agreement with very accurate simulations performed by P.
Grassberger \cite{GR2}.

Along the same lines it is possible to compute also the dynamical exponent $z$,
and then from scaling laws \cite{clar} the remaining critical exponents that
characterize the FFM model \cite{LVZ} .
Our approach can be naturally extended to the one dimensional forest fire.
We can follow the scheme used previously, and along the same lines it is
possible to compute the exponent  $\nu=1.0$ and $\tau=1.0$ which recover the
exact
results for the one dimensional case obtained in \cite{DS2}. We summarize
our results for the one and two dimensional case in Table 1.
Details of the one dimensional calculation as well as
the two dimensional case will be reported in Ref.\cite{LVZ}.

In conclusion, we introduce a novel RG approach suited to study
the critical state of the Forest-Fire model.
We identify the structure of the phase space in which  the RG transformation
is constructed and then we introduce the coupling of the renormalization
equations with the asymptotic stationarity condition of the systems.
By studying the evolution of the system under scale change,
we stress the existence of a relevant parameter, $\theta$, namely
corresponding to a repulsive fixed point for the RG equations.
In this sense, $\theta$ is the critical parameter of the model, and the
critical state is reached only with a fine tuning of $\theta$ to its
critical value.
The existence of this parameter, given by the ratio between the driving rate of
the system $f$ and the trees growth rate $p$, places the FFM  in
the field of ordinary critical phenomena. In fact, it is worth to stress that
in sandpile models the same RG analysis \cite{PVZ1} shows that no relevant
parameter is present, namely the fixed point is completely attractive,
and the slow-driving of the system does not affect the infinite correlation
properties but only the lower cut-off of distributions.

It is a pleasure to thank B. Drossel and P.Bak for interesting discussions.

\begin{table}
\centering
\begin{tabular}{|l|c|c|c|} \hline \hline
{\,\,$d=1$ \,\,}  & {\,\, $\nu$ \,\,} & {\,\, $z$ \,\,} &
{\,\,$\tau$ \,\,}\\ \hline
RG Scheme & $1.0$ & $1.0$ & $1.0$\\
Exact results$ ^{+}$ & $1.0$ & $1.0$ & $1.0$\\ \hline \hline
{$d=2$} & & & \\ \hline
RG Scheme ($2 \times 2$) & $0.73$ & $1.17$ & $1.16$\\
RG Scheme ($3 \times 3$) & $0.65$ & $1.02$ & $1.16$\\
Simulations$ ^{*}$ & $0.58$ & $1.04$ & $1.15$\\ \hline
\end{tabular}
\end{table}

\centering
In this table we summarize our results for the
critical exponents compared with exact$ ^{+}$
\cite{DS2} or experimental$ ^{*}$ results
\cite{GR2,clar}.

{\bf \large FIGURE CAPTIONS}
\begin{itemize}
\item{{\bf Fig.1:}} Configurations of sites at scale $b^{(k)}$
corresponding to a green cell at scale $b^{(k+1)}$.
\item{{\bf Fig.2:}} Example of two processes contributing to the
renormalization
of the lightning probability $f^{(k+1)}$.
\item{{\bf Fig.3:}} Example of  renormalization of the growth parameter
$p^{(k+1)}$.
\end{itemize}

\end{multicols}

\begin{thebibliography}{99}

\bibitem{BAK1} P.Bak, K. Chen and C. Tang, Phys. Lett. A {\bf 147}, 297
(1990).

\bibitem{DS} B. Drossel and F. Schwabl, Phys. Rev. Lett.,{\bf 69}, 1629
(1992).

\bibitem{bak} P.Bak, C. Tang and K. Wiesenfeld: Phys.Rev. Lett. {\bf 59},
381 (1987);  Phys. Rev. A {\bf 38}, 364 (1988).

\bibitem{GR1} P. Grassberger and H. Kantz, J. Stat. Phys.{\bf 63}, 685
(1991); W.K. Mo\ss ner, B. Drossel and F. Schwabl, Physica A {\bf 190},
205-217 (1992).

\bibitem{CLP} R. Cafiero, V. Loreto, L. Pietronero, A. Vespignani and
S. Zapperi, Europhys. Lett., {\bf 29}, 111 (1995).

\bibitem{GR2} P. Grassberger, J. Phys. A Math. Gen.{\bf 26}, 2081-2089
(1993).

\bibitem{fst} for a review see: Erzan A., Pietronero L. and Vespignani A.:
"{\em The Fixed Scale Transformation approach to fractal growth}" subm.
to Rev. Mod. Phys.; L.Pietronero, A.Erzan and C. Evertsz, Phys. Rev. Lett.
{\bf 61}, 861 (1988); R. Cafiero, L. Pietronero and A. Vespignani:
Phys. Rev. Lett.{\bf 70}, 3939 (1993);

\bibitem{PVZ1} L.Pietronero, A. Vespignani and S. Zapperi, Phys. Rev. Lett.
{\bf 72}, 1690 (1994); A.Vespignani, S.Zapperi and L.Pietronero,
Phys. Rev. E {\bf 51}, 1711 (1995).

\bibitem{DS1} B. Drossel and F. Schwabl, Physica A {\bf 204},
212-229 (1994).

\bibitem{CH}
K. Christensen, H. Flyvbjerg and Z.Olami,
Phys. Rev. Lett. {\bf 71}, 2737 (1993).

\bibitem{clar}
S.Clar, B.Drossel and F.Schwabl, Phys. Rev. E {\bf 50}, 1009 (1994).

\bibitem{LVZ} V. Loreto, A. Vespignani and S. Zapperi,
"Renormalization Scheme for Forest-Fire models", submitted to J. of Phys.

\bibitem{DS2} B. Drossel, S. Clar and F Schwabl, Phys. Rev. Lett. {\bf 71},
3739 (1993);

\end{thebibliography}
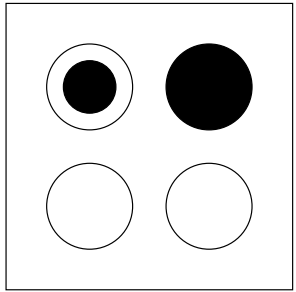
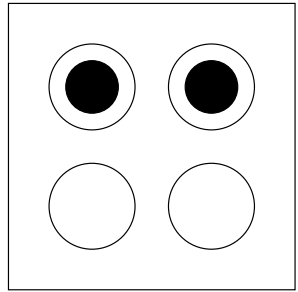
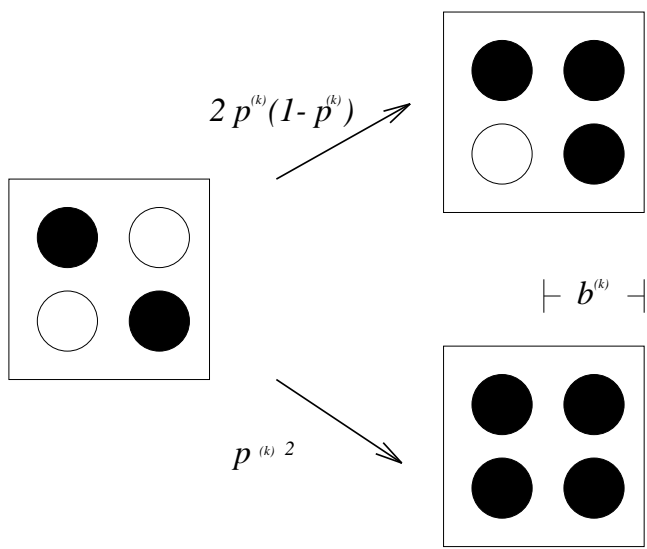
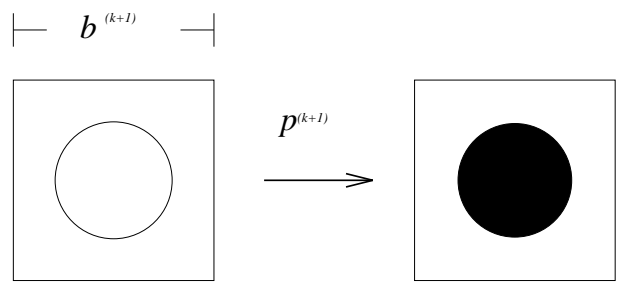
\end{document}